# Interactions in an acoustic world

Ion Simaciu[1,a], Gheorghe Dumitrescu[2,b], Zoltan Borsos[1] and Mariana Brădac[1]

[1] Petroleum-Gas University of Ploieşti, Ploieşti 100680, Romania

[2] High School Toma N. Socolescu, Ploieşti, Romania

## Abstract

*The present paper aims to complete an earlier paper where the acoustic world was introduced. This is accomplished by analyzing the interactions which occur between the inhomogeneities of the acoustic medium, which are induced by the acoustic vibrations traveling in the medium. When a wave packet travels in a medium, the medium becomes inhomogeneous. The spherical wave packet behaves like an acoustic spherical lens for the acoustic plane waves. According to the principle of causality, there is an interaction between the wave and plane wave packet. In specific conditions the wave packet behaves as an acoustic black hole.*

## 1. Introduction

Acoustic waves carry energy, momentum, angular momentum and also physical information regarding the source and the medium through which they propagate. Therefore one can say that the acoustic waves intermediate interactions occurring within the medium. Nowadays extended theoretical [1-14] and experimental [15-17] research work is being carried out to study the interactions induced by acoustic waves.

One can distinguish two types of interactions induced by acoustic waves. The former type are those which are the interactions between solids, liquids and gases (cavities, bubbles, drops) immersed into fluid and contained into a solid.

The latter are those interactions which occur between the inhomogeneities of the medium and which are induced by acoustic waves.

The first type of interactions mentioned above were early studied by de Basset, A. B. [1]. and substantiated by Bjerknes, C. A. and Bjerknes, V. F. K. [2, 3].

In what it follows we will assume that the acoustic processes are similar to the processes occurring into vacuum. That is why we will study those processes occurring within a medium, produced by acoustic perturbations. When waves interfere they generate wave packets. These wave packets behave like the inhomogeneities of the medium. Also these inhomogeneities induce new inhomogeneities due to the combined process diffraction-interference-absorption.

Cavities may appear in the core of the wave packets [18] when the energy of the oscillations is of the same order as the binding energy of the particles of the liquid. All these extended

[a] sion@upg-ploiesti.ro; isimaciu@yahoo.com

[b] meditatie@yahoo.com

inhomogeneities behave as sources for oscillations and as diffracting objects.

The container where the fluid is, or the boundaries of the solid, play an important role for these processes. The container behave like a cavity which reflects the acoustic waves inwards. We name an acoustic world the finite volume of the fluid or, of the solid, together with the processes occurring within [19]. Within the acoustic world the microscopic parameters of the inhomogeneities and the container parameters are related. For this reason, the microscopic parameters of the inhomogenities and of the enclosure are dependent on each other. Their relationship is similar to the one between the parameters of the elementary particles and those of the Universe [20], according to the present approaches.

In this paper we aim to study which features of the interactions may be generated by an inhomogeneity which is induced by a wave or a wave packet.

In the second section of the paper we infer the expression of the wave propagation velocity in a fluid when the density and the pressure vary adiabatically. Using this expression, we infer the acoustic expression of the refractive index.

In the third section we deduce the expression of the acoustic index of refraction of a fluid perturbed by a propagating wave and a standing wave.

In the fourth section we achieve the expression of the acoustic index of refraction of a fluid perturbed by a stationary wave packet. The averaged refractive index depends on the position related to the center of the wave packet.

In the fifth section we argue that a wave which propagates into a fluid, whose properties were changed by a wave packet, deviates from its initial direction of propagation. Also we analyze the possible existence of the acoustic black hole. The radius of the acoustic black hole depends on the wavelength and of the oscillation amplitude of the acoustic wave packet.

Sixth section is dedicated to conclusions.

## 2. Phase velocity in the presence of an inhomogeneity

### 2.1. Phase velocity

In a prior papers [19] we have studied the issue of the dual behaviour of matter related to the longitudinal waves. In what it follows we will study how these waves may change the density of the medium through which they travel. According to Landau and Lifchitz [21, Ch.8, §63], the propagation of a mechanical wave through a medium changes the local density $\rho = \rho_0 + \delta\rho$ the pressure $p = p_0 + \delta p$ and the temperature $T = T_0 + \delta T$.

The phase velocity in a fluid is

$$u = \sqrt{\left(\frac{\partial p}{\partial \rho}\right)_S} \tag{1}$$

In the case of an ideal gas, the process of the wave propagation is adiabatic and the pressure is related to the density according to the relation

$$p = p_0 \left( \frac{\rho}{\rho_0} \right)^\gamma \qquad (2a)$$

and

$$\rho = \rho_0 \left( \frac{p}{p_0} \right)^{\frac{1}{\gamma}}. \qquad (2b)$$

Using (2) into (1) we can find the velocity of a wave travelling in modified medium, due to the presence of an adiabatic perturbation, as

$$u = u_0 \sqrt{\left( \frac{\rho}{\rho_0} \right)^{\gamma-1}} = u_0 \left( 1 + \frac{\delta\rho}{\rho_0} \right)^{\frac{\gamma-1}{2}}. \qquad (3a)$$

or

$$u = u_0 \sqrt{\left( \frac{p}{p_0} \right)^{\frac{\gamma-1}{\gamma}}} = u_0 \left( 1 + \frac{\delta p}{\rho_0} \right)^{\frac{\gamma-1}{2\gamma}}. \qquad (3b)$$

For a compressible liquid [22, p. 44] the dependency between the phase velocity and the density is

$$u = u_0 \left( \frac{\rho}{\rho_0} \right)^{\frac{\beta-1}{2}} = u_0 \left( 1 + \frac{\delta\rho}{\rho_0} \right)^{\frac{\beta-1}{2}}, \qquad (4a)$$

or

$$u = u_0 \left( \frac{p}{p_0} \right)^{\frac{\beta-1}{2\beta}} = u_0 \left( 1 + \frac{\delta p}{p_0} \right)^{\frac{\beta-1}{2\beta}}, \qquad (4b)$$

with $\beta$, constant which depend on the liquid type.

One notices the difference between these two expressions of velocities, (3) and (4), due to the difference $\beta \neq \gamma$. One can generalize the phase velocity of a longitudinal wave for all states of the matter

$$u = u_0 \left( 1 + \frac{\delta\rho}{\rho_0} \right)^\zeta. \qquad (5a)$$

or

$$u = u_0 \left( 1 + \frac{\delta p}{p_0} \right)^{\frac{\zeta}{2\zeta+1}} \qquad (5a)$$

For a gaseous state of the matter $\zeta = (\gamma-1)/2$ is subunitary and for a liquid state of the matter, it is greater than 1 (e.g., for water $\beta \cong 7$ and hence $\zeta = (\beta-1)/2 \cong 3$ [22, p. 44].

According to the relations (3-5), for every state of the matter, the increasing of the density ($\delta\rho > 0$), due to the deformation, implies a growth of the velocity. Conversely, the decreasing of the density ($\delta\rho < 0$) leads to a decrease in velocity.

## 2.2. The acoustic index of refraction

We will assign an acoustic index of refraction to the perturbed medium as

$$n_a = \frac{u_0}{u}. \tag{6}$$

According to (5a) and with $\delta\rho \ll \rho_0$, the above definition leads to

$$n_a(\rho) = \left(1 + \frac{\delta\rho}{\rho_0}\right)^{-\zeta} \cong 1 - \zeta\frac{\delta\rho}{\rho_0} + \frac{\zeta(\zeta+1)}{2}\left(\frac{\delta\rho}{\rho_0}\right)^2 \tag{7a}$$

and, according to (5b), with $\delta p \ll p_0$

$$n_a(p) = \left(1 + \frac{\delta p}{p_0}\right)^{\frac{-\zeta}{2\zeta+1}} \cong 1 - \frac{\zeta}{2\zeta+1}\frac{\delta p}{p_0} + \frac{\zeta(3\zeta+1)}{2(2\zeta+1)^2}\left(\frac{\delta p}{p_0}\right)^2 \tag{7b}$$

The averaged expressions of the acoustic index of refraction are:

$$\langle n_a(\rho)\rangle \cong 1 - \zeta\frac{\langle\delta\rho\rangle}{\rho_0} + \frac{\zeta(\zeta+1)}{2}\frac{\langle(\delta\rho)^2\rangle}{\rho_0^2} \tag{8a}$$

$$\langle n_a(p)\rangle \cong 1 - \frac{\zeta}{2\zeta+1}\frac{\langle\delta p\rangle}{p_0} + \frac{\zeta(3\zeta+1)}{2(2\zeta+1)^2}\frac{\langle(\delta p)^2\rangle}{p_0^2} \tag{8b}$$

According to the results of [21, ch.8, §64], there are the relations:

$$\langle\delta\rho\rangle = \frac{\langle(\delta p)^2\rangle}{2}\left(\frac{\partial^2\rho}{\partial p^2}\right)_{S,p=p_0} = -\zeta\rho_0\frac{\langle v^2\rangle}{u_0^2}, \tag{9a}$$

$$\langle\delta p\rangle = 0, \tag{9b}$$

$$\langle(\delta\rho)^2\rangle = \rho_0^2\frac{\langle v^2\rangle}{c_0^2}, \tag{9c}$$

$$\langle(\delta p)^2\rangle = \rho_0^2 c_0^2 \langle v^2\rangle. \tag{9d}$$

By replacing the relations (9) in the expression of the averaged acoustic index of refraction (8b) yields:

$$\langle n_a(\rho)\rangle \cong 1 + \frac{\zeta(3\zeta+1)}{2}\frac{\langle v^2\rangle}{u_0^2} \tag{10a}$$

The same result is obtained by replacing relations (9) in the expression of the averaged acoustic index of refraction (8a)

$$\langle n_a(p) \rangle \cong 1 + \frac{\zeta(3\zeta+1)}{2(2\zeta+1)^2} \frac{\langle (\delta p)^2 \rangle}{p_0^2} = \langle n_a(\rho) \rangle \tag{10b}$$

Hence, we find that the index of refraction is a measure of the inhomogeneity of the medium.

## 3. The inhomogeneity induced by a wave

According to [21, ch.8, §63], the relative variance of the density, in the presence of a wave, is the ratio between the velocity of the vibration $\dot{q}(x,t)$ and the phase velocity $u_0$

$$\frac{\delta \rho}{\rho} = \frac{\upsilon}{u_0} = \frac{\dot{q}}{u_0} \ll 1 \tag{11}$$

Substituting (11) in (10a), with $\gamma p_0 = (2\zeta+1) p_0 = \rho_0 u_0^2$, the expression of the index of refraction becomes

$$\langle n_a \rangle \cong 1 + \frac{\zeta(3\zeta+1)}{2} \frac{\langle \dot{q}^2 \rangle}{u_0^2} \tag{12}$$

For a plane wave $\Psi(z,t) = q_z = q_{0z} \sin(\omega t - kz)$, the velocity of the oscillation is

$$\dot{q}_z = q_{0z} \omega \cos(\omega t - kz), \tag{13}$$

Averaging, in time, the square velocity, one obtains

$$\langle \dot{q}_z^2 \rangle_t = \frac{q_{0z}^2 \omega}{T} \int_0^T \cos^2(\omega t - kz) d(\omega t) = \frac{q_{0z}^2 \omega^2}{2}, \tag{14a}$$

$$\frac{\langle \dot{q}_z^2 \rangle_t}{u_0^2} = \frac{q_{0z}^2 \omega^2}{2 u_0^2} = \frac{2\pi^2 q_{0z}^2}{\lambda_0^2} \ll 1 \text{ or } q_{0z} \ll \lambda. \tag{14b}$$

Substituting (14) in (12), the expression of the index of refraction becomes

$$\langle n_a \rangle_t \cong \left[ 1 + \frac{\zeta(3\zeta+1)}{2} \left( \frac{q_{0z} \omega}{u_0} \right)^2 \right] > 1. \tag{15}$$

Hence for a standing wave, $\Psi_s(z,t) = q_{zs} = q_{0zs} \sin(kz) \sin(\omega t)$ [31], with the velocity of oscillation

$$\dot{q}_{zs} = q_{0zs} \omega \sin(kz) \cos(\omega t). \tag{16}$$

The averaged velocity of oscillation through time is zero, but the averaged square velocity of oscillation through time is not zero.

$$\langle \dot{q}_{zs}^2 \rangle_t = \frac{q_{0zs}^2 \omega \sin^2 kz}{T} \int_0^T \cos^2(\omega t) d(\omega t) = \frac{q_{0zs}^2 \omega^2}{2} \sin^2 kz. \tag{17}$$

Now, according to (15), the following expressions occurs:

$$\langle n_a \rangle_t \cong \left[ 1 + \frac{\zeta(3\zeta+1)}{4} \left( \frac{q_{0zs}\omega}{u_0} \sin kz \right)^2 \right] \tag{18}$$

which depicts the dependence of the index to the position $z$. Also the process mentioned above involves the change of the optical index, due to the changes of the mass density of the medium. This phenomenon is used as a diffraction grating [23, 24].

Performing a spatial average $\langle \sin^2 kz \rangle_z = 1/2$, yields

$$\langle n_a \rangle_{ts} \cong 1 + \frac{\zeta(3\zeta+1)}{8} \left( \frac{q_{0zs}\omega}{u_0} \right)^2 = 1 + \frac{\zeta(3\zeta+1)}{2} \left( \frac{q_{0z}\omega}{u_0} \right)^2, \tag{19}$$

since the amplitude of the standing wave is double of the amplitude of the non standing wave $q_{0zs} = 2q_{0z}$. The relations (15, 18 and 19) display how a longitudinal wave changes the acoustic index of a medium.

## 4. Inhomogeneity induced by a wave packet

### 4.1. The acoustic index of refraction in the presence of a wave packet

If we assume a three dimensional standing wave packet $\Psi_p(r,t) = q_p = q_{0p}[\sin(k_0 r)/k_0 r]\sin(\omega_0 t)$. Its velocity of oscillation is

$$\dot{q}_p = q_{0p}\omega_0 \frac{\sin(k_0 r)}{k_0 r} \cos(\omega_0 t) = u_0 q_{0p} \frac{\sin(k_0 r)}{r} \cos(\omega_0 t). \tag{20}$$

We will assume that the medium becomes spatially and temporally inhomogeneous. Its temporal averaged velocity of oscillation is null. But its averaged square velocity is not

$$\langle \dot{q}_p^2 \rangle_t = \frac{q_{0p}^2 \omega_0^2 \sin^2 k_0 r}{(k_0 r)^2 T_0} \int_0^{T_0} \cos^2(\omega_0 t) d(\omega_0 t) = \frac{q_{0p}^2 \omega_0^2}{2} \frac{\sin^2 k_0 r}{(k_0 r)^2}. \tag{21}$$

Now we can express the temporal averaged acoustic index as

$$\langle n_a \rangle_t \cong \left[ 1 + \frac{\zeta(3\zeta+1)}{4} \left( q_{0p} \frac{\sin k_0 r}{r} \right)^2 \right] > 1. \tag{22}$$

Hence, according to (22), the index of refraction depends on the position ($r$).

## 5. Acoustic black hole

### 5.1. The deflection of the acoustic waves

Using the above index of refraction one can infer the deflection of a wave which travels through a medium, when the medium is modified by the wave packet. A simple calculus [25, 26] one can perform if one assumes an analogy between the medium with variable index and a spherical lens, with radius of curvature $R_1 = -R_2 = r$ and the index of refraction from (22). The focal length of the acoustic lens is [27, ch. 27.3]

$$\frac{1}{f} = \frac{2}{r}(\langle n_a \rangle_t - 1) = \frac{\zeta(3\zeta+1)}{2r}\left(q_{0p}\frac{\sin k_0 r}{r}\right)^2 > 0. \tag{23}$$

Hence, the lens is a converging lens. For small angles, the deflection of the wave (which far away from the lens is null and where $r$ is the impact parameter) is

$$\varphi \cong \tan\varphi = \frac{r}{f} = 2(\langle n \rangle_{trm} - 1) = \frac{\zeta(3\zeta+1)}{2}\left(q_{0p}\frac{\sin k_0 r}{r}\right)^2. \tag{24}$$

From (22) one can see that the wave deviates from initial direction and this deflection can be interpreted as an attraction of the wave by the wave packet. This interaction is similar to the interaction between light and a body with gravitational mass.

## 5.2. The radius of acoustic horizon

The wave packet behaves like an acoustic black hole (dumb hole, sonic black hole) where the focal length of the lens is equal to the acoustic impact parameter

$$f = r = r_a. \tag{25}$$

The acoustic hole corresponding to the acoustic wave packet is static. This hole is different from the acoustic hole or acoustic wormhole generated by fluid flow [28-30].

Replacing the focal length (25) into the expression (23), we obtain the equation for the radius of the acoustic black hole or the radius of the acoustic horizon

$$\frac{\sin k_0 r_a}{k_0 r_a} = \frac{1}{k_0 q_{0p}}\sqrt{\frac{2}{\zeta(3\zeta+1)}}. \tag{26}$$

For small angles $k_0 r_a < \pi/36$, with the approximation $\sin k_0 r_a \cong k_0 r_a\left[1 - (k_0 r_a)^2/6\right]$, the physical solution is

$$r_a = \frac{6}{k_0}\sqrt{1 - \frac{1}{k_0 q_{0p}}\sqrt{\frac{2}{\zeta(3\zeta+1)}}} = \frac{3\lambda_0}{\pi}\sqrt{1 - \frac{\lambda_0}{q_{0p}}\frac{1}{\pi\sqrt{2\zeta(3\zeta+1)}}}. \tag{27}$$

The solution given by (27) is real if

$$\frac{\lambda_0}{q_{0p}} < \pi\sqrt{2\zeta(3\zeta+1)}. \tag{28}$$

For the liquid, with $\zeta \cong 3$, yields

$$\lambda_0/q_{0p} < \sqrt{60}\pi. \tag{29}$$

If we take into account the relationship between the wave packet parameters and the fluid parameters [19a, §3.1; 19b, §5.3]:

$$q_{0p} < a, \quad \lambda_0 \geq 2a, \tag{30}$$

it follows that

$$\frac{a}{\pi\sqrt{15}} < q_{0p} < a. \tag{31}$$

Those restrictive requirements addressed to the oscillation amplitude and the wavelength are related to the binding energy of the liquid particles (molecules). When the wavelength and the amplitude meet the boundaries one generates a bubble in the center of the wave packet. It follows that the acoustic black hole (dumb hole) is associated with the generation of a bubble. The bubbles are able to interact each other in the presence of the acoustic waves, and therefore in the presence of the waves background of the acoustics world. Then a further development of the papers [1-18] can be done starting from this issue.

The restrictive requirement (14b), is in contradiction with the requirement (29). In order to achieve an appropriate study of the existence of the acoustic black hole we have to address to the issue of the acoustic refractive index related to high density variation, i.e. $\delta\rho \leq \rho_0$ ).

## 6. Conclussions

We have proved that a fluid becomes inhomogeneous in interaction with a wave or a wave packet. The inhomogeneity involves the change of the propagation velocity of a plane wave. The change of the speed of propagation of a wave is similar to the existence of an acoustic refractive index.

For a spherical wave packet, the refractive index depends on the position from the center of the wave packet. The spherical wave packet induces a change in the trajectory of the acoustic plane wave. Following the approach of the gravitational deviation of the light we computed the deviation of a plane wave in the presence of the wave packet. According to the principle of causality, an interaction may obey between the wave and the plane wave packet.

The deviation depends on the wave packet parameters: the wavelength and the amplitude of oscillations and the position relative to the center of the wave packet. The existence of the deviation of the plane wave traveling in the fluid (modified for spherical wave packet) suggests that for certain values of the impact parameter, the plane wave has a circular trajectory. This is an acoustic black hole.

In our approach we have adopted the same frequency for the plane wave as for the waves of the wave packets which disturb the fluid. In a forthcoming paper we will analyze the consequences which can arise from the assumption that the plane wave has another angular frequency and therefore the average is done for a period corresponding to this angular frequency.